\documentclass{nature}

\usepackage[british,UKenglish]{babel}

\usepackage{amsmath,amssymb}
\usepackage[caption=false]{subfig}
\usepackage{color}
\usepackage{xcolor}
\usepackage{rotating}
\usepackage{bm}
\usepackage{hyperref}
\usepackage{float}
\usepackage{soul}
\usepackage{color}
\usepackage{braket}
\usepackage{physics}
\usepackage{bbold}
\usepackage{nicefrac}

\renewcommand{\vec}[1]{\bm{#1}}

\usepackage{graphicx}
\makeatletter
\let\saved@includegraphics\includegraphics
\AtBeginDocument{\let\includegraphics\saved@includegraphics}
\renewenvironment*{figure}{\@float{figure}}{\end@float}
\makeatother

\title{Coexistence of ultraheavy and ultrarelativistic Dirac quasiparticles in sandwiched trilayer graphene}

\author{Stephen Carr,$^{1,*}$ Chenyuan Li,$^{1,*}$
Ziyan Zhu,$^{1}$
Efthimios Kaxiras,$^{1,2}$
Subir Sachdev$^1$
\& Alex Kruchkov$^{1}$ 
 }

\begin{document}

\maketitle

\begin{affiliations}
 \item Department of Physics, Harvard University, Cambridge, MA 02138, USA
 \item Harvard J.A. Paulson School of Engineering and Applied Sciences,  Cambridge, MA 02138, USA
\end{affiliations}

\begin{abstract}
Electrons in quantum materials exhibiting coexistence of dispersionless (flat) bands piercing dispersive (steep) bands can give rise to strongly correlated phenomena, and are associated with unconventional superconductivity. It is known that   in twisted trilayer graphene steep Dirac cones can coexist with band flattening, but the phenomenon is not stable under layer misalignments.
Here we show that such a twisted sandwiched graphene (TSWG) -- a three-layer van der Waals heterostructure with a twisted middle layer -- can have very stable flat bands coexisting with Dirac cones near the Fermi energy when twisted to 1.5$^{\circ}$.
These flat bands require a specific high-symmetry stacking order, and 
our atomistic calculations predict that TSWG always relaxes to it. 
Additionally, with external fields, we can control the relative energy offset between the Dirac cone vertex and the flat bands. 
Our work establishes twisted sandwiched graphene as a new platform for research into strongly interacting phases, and topological transport beyond Dirac and Weyl semimetals.
\end{abstract}

Graphene, an atomically thin crystal of carbon, provides an experimentally favorable platform for two dimensional (2D) Dirac physics as it exhibits ultrarelativistic Dirac cones in its band structure,  described with massless quasiparticles when weak spin-orbit coupling is neglected.\cite{CastroNeto2009}
Bilayers of graphene in the energetically favorable Bernal ($AB$) stacking have quadratic dispersion and quasiparticles with well-defined effective mass.\cite{CastroNeto2009} 
Twisted bilayer graphene (TBG) -- two rotationally mismatched graphene layers -- can be fabricated at the so-called \textit{magic angle} near $1.1^{\circ}$, where it hosts ultraheavy fermions with remarkably flat, almost dispersionless electronic bands\cite{Trambly2010,Bistritzer2010,Morell2010,Tarnopolsky2018} of a topological origin.\cite{Tarnopolsky2018,Liu2019a,Lian2018,Kang2018}
The twist angle serves as a precise control of the interlayer coupling between the graphene monolayers, revealing the band flattening  phenomena as an ultimate manifestation of hybridization of Dirac cones.
Flat bands and the corresponding large density of electronic states can lead to novel strongly correlated phenomena.
Indeed, since the recent discovery of correlated insulators and unconventional superconductivity in TBG,\cite{Cao2018a,Cao2018b,Yankowitz2018,Lu2019Efetov} van der Waals multilayer stacks have been further explored as a platform of exotic correlated physics. In particular, effectively 2D heterostructures consisting of flat sheets of graphene, transition metal dichalcogenides, and hexagonal boron nitride have been successful candidates for the moir{\'e}-induced correlated phenomena.\cite{Chen2019,Zhang2018chern,Wu2019,Naik2018,Alexeev2019, Zhang2018,Chebrolu2019,Koshino2019,Haddadi2019,Zhao2019,Xian2018multi} Recent experimental progress in studying correlations in multilayer heterostructures with more than two twisted graphene layers\cite{Chen2019,Liu2019,Cao2019,Shen2019}  has led to a search for novel multilayer platforms with a particular focus on the trilayer geometry.\cite{Amorim2018,Zhang2018chern,Khalaf2019, Mora2019,Ma2019}

In this work, we provide a detailed \textit{ab initio} study of a unique extension of the TBG system: the twisted graphene sandwich (Fig. \ref{fig:bands_fig}b), which is a promising construct of a three-layer graphene  heterostructure.\cite{Khalaf2019} In general, different trilayer systems are also represented by the untwisted $ABC$ stack, the twisted monolayer on bilayer, and the doubly incommensurate twisted trilayer.
The $ABC$ graphene stack has been well understood,\cite{CastroNeto2009} although it has recently been observed to host correlated states when placed on hexagonal boron nitride due to a lattice-mismatch induced moir\'e superlattice.\cite{Chen2019}
The twisted monolayer on bilayer system is of the same experimental complexity as the twisted bilayer, and can host both parabolic and Dirac cone bands near the Fermi energy,\cite{Morell2013,Zhang2018chern} but has a less robust flat band at its magic angle regime due to its reduced symmetry (Fig. \ref{fig:bands_fig}c).
The doubly incommensurate twisted trilayer is challenging to model accurately due to a complicated umklapp scattering process mediated by two independent twist angles, and the existing studies do not show the same spectacular flat bands as in the twisted bilayer.\cite{Mora2019}

In contrast, the graphene sandwich retains a high degree of symmetry.\cite{Khalaf2019}
In the meantime, it has an effectively stronger interlayer coupling between layers, promising flat bands at larger angles and thus smaller moir\'e length-scales, likely enhancing correlated effects. We show in this work that the trilayer system hosts a unique feature compared to the twisted bilayer: a symmetry-protected Dirac cone that pierces through the magic-angle flat band. However, the system poses an experimental challenge: to perfectly mimic the moir\'e pattern of TBG, the bottom and top layers of TSWG need to be aligned in $AA$ stacking. 
We show that this challenge is overcome by natural relaxation of the sandwiched heterostructure, leading to protected coexistence of ultraheavy and ultrarelativistic Dirac quasiparticles at the same energy scale.

\section*{Results}

\begin{figure}
\centering{
\includegraphics[width=0.7  \linewidth]{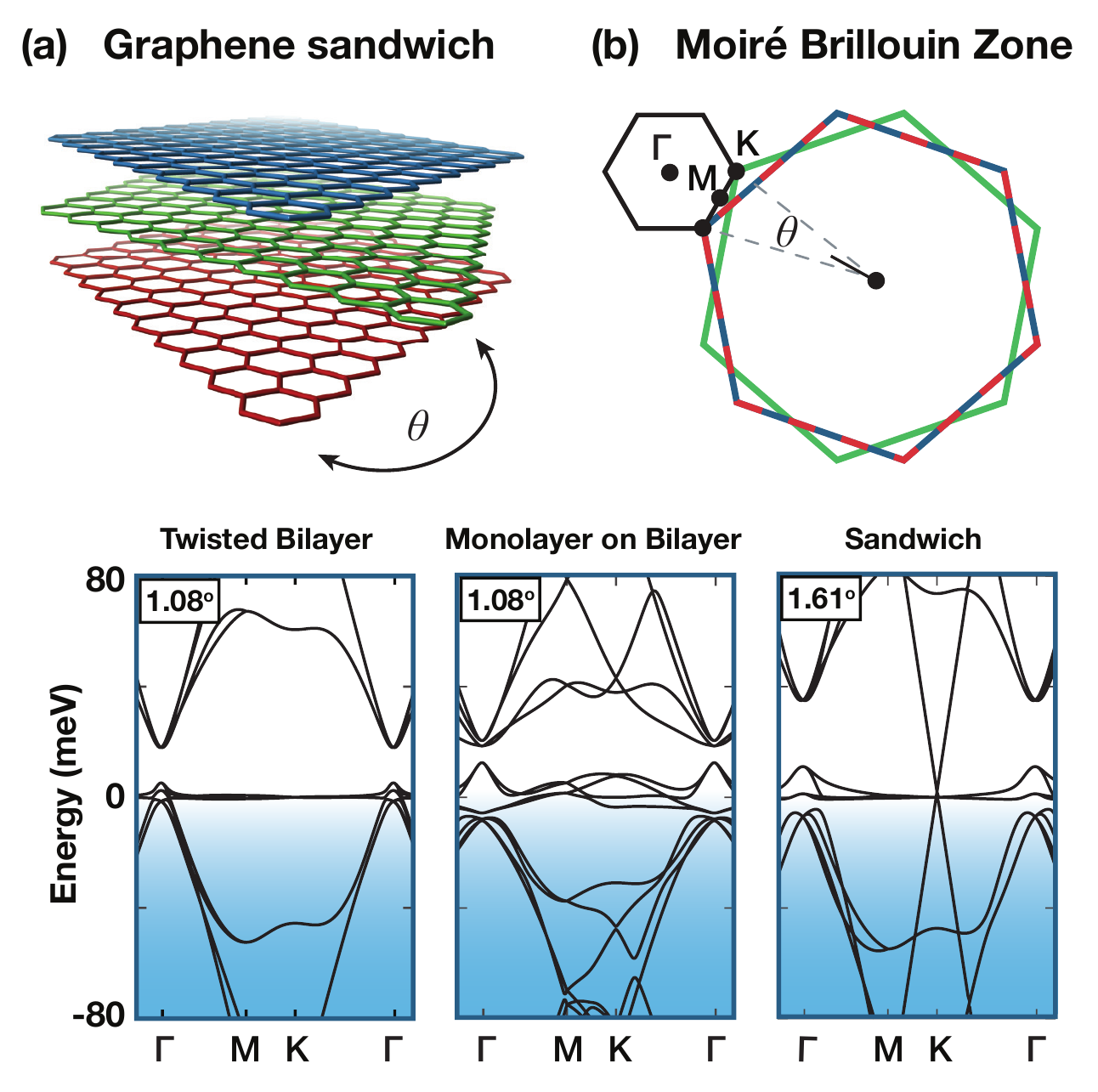}}
	\caption{ \footnotesize
	\textbf{Twisted sandwich graphene and
	\textit{ab initio} tight-binding band structures for twisted graphene stacks.}
	\textbf{(a)} Schematic of the graphene sandwich: the middle layer is rotated  by $\theta$, while the "bread" layers are aligned.
	\textbf{(b)}  
	For the sandwiched graphene, the Brillouin zones of the bread layers are identical (red\&blue striped line); the resulting  moir\'e Brillouin zone (mBZ) is depicted in black.
	\textbf{(c)} Comparison for (unrelaxed) band structures with the same single-twist mBZ:  (left) Band structure of twisted bilayer graphene (TBG) at the magic angle 1.08$^\circ$; (center): Band structure of monolayer graphene twisted on bilayer AB graphene (MG/BG) at the same angle; (right) Band structure of the twisted sandwiched graphene (TSWG) at it's (unrelaxed) magic angle $\theta$=1.61$^\circ$. Already in unrelaxed atomistic calculations, the TSWG reveals a remarkable coexistence of Dirac cones pierced by ultraflat bands.
	\label{fig:bands_fig}
	}
\end{figure}

\begin{figure}
\vspace{-10 mm}
\centering{
\includegraphics[width=1 \linewidth]{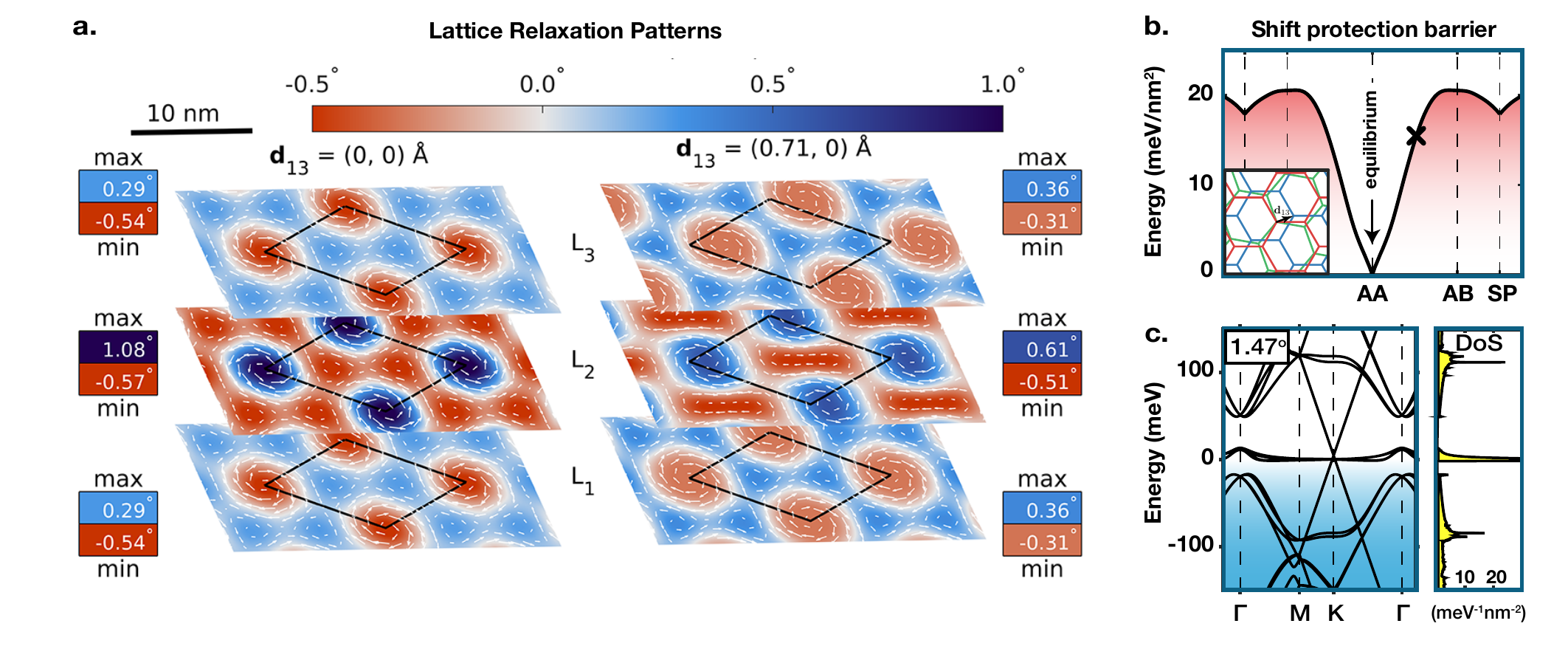}}
\vspace{-15 mm}
			\caption{
			\footnotesize
			\textbf{Energetic stability and electronic effects of lattice relaxations on the flat bands.}
	 \textbf{(a)} Atomic relaxations in the graphene sandwich with $1.47^\circ$ twisting angle, obtained via a extended continuum model.
Atomic displacements in each layer ($\mathbf{u}(\vec r)$) are visualized with white arrows (not to scale); the color data denotes information on the local value of the in-plane twisting ($\Delta \theta$) due to the relaxations ($\nabla \times \mathbf{u}$), with positive $\Delta \theta$ corresponding to counter-clockwise rotation.
	The moir\'e supercell is outlined in black.
	When the relative stacking between layer $1$ and $3$ ($\mathbf{d}_{13}$) is unconstrained, the system always reaches minimum energy by translating back to $\mathbf{d}_{13} = 0$.
	The relaxation when $\mathbf{d}_{13}$ is nonzero is weaker, and the overall energy is higher compared to the unconstrained case.
	\textbf{(b)} Energy as a function of the stacking configuration $\mathbf{d}_{13}$, between Layers 1 and 3, with the high-symmetry stackings highlighted ($AA$, $AB$, and Saddle Point).
	The black `x' indicates the stacking shown in the second relaxation plot ($\mathbf{d}_{13} \neq 0$) on subfigure \textbf{(a)}. We see that the AA stacking of bread layers, vital for coexistence of flat bands with Dirac cones, is protected by a large energy barrier of approximately 20 meV/nm$^2$. Inset figure: Diagram of $\mathbf{d}_{13}$, defined as the vector displacement between the $A$ orbital of $L_1$ and the $A$ orbital of $L_3$.
	\textbf{(c)} Fully relaxed TSWG band structure (tight-binding calculations) at the redefined magic angle $1.47^\circ$  and  the corresponding density of states (right panel). Protected by relaxations towards AA stacking, the flat bands and Dirac cones coexist at the same energy scale and are slightly offset by just $4$ meV. \label{fig:relaxations}}
\end{figure}

\textbf{Electronic structure.}---
The electronic states of the twisted graphene sandwich consist of two main features near the Fermi energy: a set of four nearly flat bands, similar to those found in twisted bilayer graphene, and a Dirac cone reminiscent of monolayer graphene.
Much like twisted bilayer graphene, the flatness of the first feature depends sensitively on the twist angle, crystal relaxation, and external perturbations.\cite{Nam2017, carr2019}
The sandwich's Dirac cone is nearly identical to that of the monolayer cone, with a Fermi velocity of $0.75 \times 10^6$ m/s compared to $0.81 \times 10^6$ m/s in the monolayer case.
We note that the sandwiched trilayer graphene has an advantage compared to the twisted monolayer on bilayer graphene (BG/MG), where the effects of band flattening are obscured by non-symmetric band hybridization due to the absence of a layer-inversion symmetry between the monolayer and bilayer materials (see Fig. \ref{fig:bands_fig}).
In contrast, the magic angle graphene sandwich shows a Dirac cone piercing nearly flat bands, which in the non-interacting picture already classifies it as an unconventional semimetal. In our \textit{ab initio} calculations, the principal magic angle is found at $1.61^{\circ}$ (see Fig. \ref{fig:bands_fig}) for a rigid system, and inclusion of realistic lattice relaxation effects sharpens it to $1.47^{\circ}$ (see Fig. \ref{fig:relaxations}). 

\begin{figure}
\centering{
\includegraphics[width=1 \linewidth]{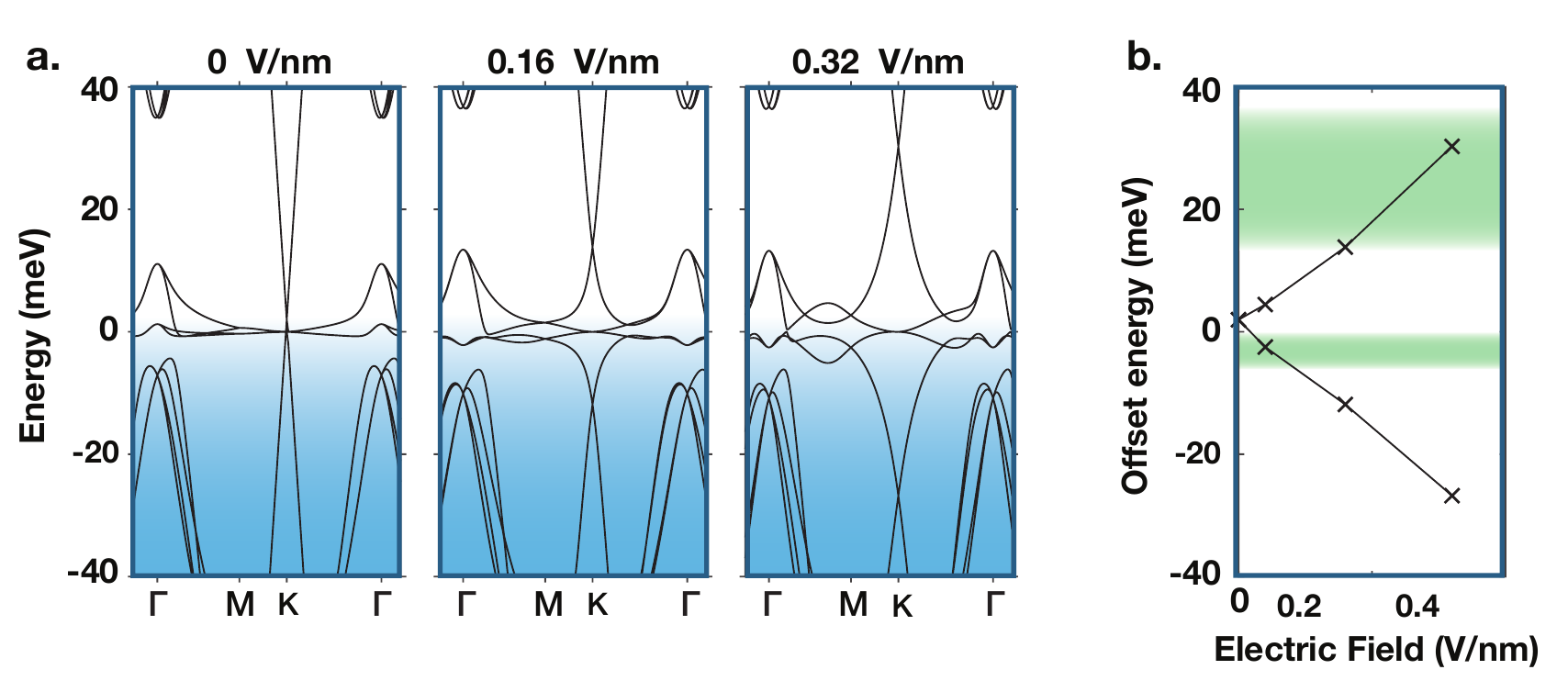}}	
	\caption{\footnotesize
	\textbf{Controlling hybridization and the Dirac cone offset with external fields.}
	\textbf{(a)}  When the electric field is non-zero, the Dirac-cone splits into two, one above and one below the flat bands,
	and the flat bands have a parabolic touching point at $K$ (unrelaxed band structures at 1.61$^{\circ}$). 
	\textbf{(b)} 
	The extracted offset energy $\Delta E_K$ as a function of electric fields is shown with the green regions representing the energy regions where the Dirac-cone will overlap with other electronic states. Clearly, the offset energy $\Delta E_K$ is linear in moderate electric fields, providing precise control of the relative position of the Dirac cone vertex and the intensity of the  van Hove singularity associated with the flattened bands.
	\label{fig:tuning_fig}}	
\end{figure}

The electronic structure of the TSWG can be explained by considering the top and bottom layers as one effective layer.
We assume that the top and bottom layers are aligned to ensure that they have the identical electronic coupling to the middle layer.
Then, the effective states are odd or even combinations between the $p_z$ orbitals of the top and bottom carbon atoms.
The even combinations can couple with the middle layer, with an interlayer coupling a factor of $\sqrt{2}$ stronger than that of the twisted bilayer graphene, moving the flat band regime from $\theta = 1.1^\circ$ to roughly $\theta = 1.5^\circ$.
The odd combination is decoupled from the middle layer, leaving a copy of the pristine monolayer Dirac cone being pierced by the flat bands.
In fact, this decoupling is not exact, and we report a weak coupling between the ``bread'' of the sandwich (top and bottom graphene layers) of $6\,$meV in density functional theory (DFT) calculations, which is ignored in our tight-binding and continuum models in this work. The weak coupling could be important for the strongly-correlated electron physics. 
If included, the coupling can cause weak hybridization between the flat bands and the Dirac cone, similar to the effect of vertical displacement fields (see Fig. \ref{fig:tuning_fig}c).

In realistic TSWG, the vertex of the Dirac cone is slightly off-set from the flat bands at the $K$ point (see Fig. \ref{fig:relaxations}).
The off-set energy $\Delta E_K$ (the energy difference between the Dirac cone vertex and the flat bands) is always small and positive, meaning the flat electronic bands are always piercing the Dirac cone slightly above the vertex in our study.
We report that the exact value of $\Delta E_K$ appears to be very sensitive to the parameterization of the model.
Throughout this work, we use a model for monolayer graphene that includes up to the third nearest-neighbour and accounts for strain effects.\cite{Fang2018}
This model gives $\Delta E_K = 2$ meV.
However, if using an older model with up to eight nearest-neighbours (but no strain corrections),\cite{Fang2016} we obtain $\Delta E_K = 10$ meV.
If we modify this model by truncating the range of the coupling, $\Delta E_K$ reduces smoothly to the $2\,$meV result at third nearest-neighbour.
To explain this strong dependence on the monolayer model, it is important to understand the origin of $\Delta E_K$.
The Dirac cone is effectively decoupled from the twisted system, and so its vertex lies at the same energy as it would in the monolayer case, at the monolayer Fermi level.
The flat bands of TSWG, however,  have a modified Fermi energy $\Delta E$ due to the interlayer coupling over the moir\'e cell.
The shift $\Delta E$ in the flat bands Fermi energy is not well documented in the existing TBG literature, as it can always be safely ignored by fixing the Fermi energy of the bilayer system to zero after a band structure calculation is performed.
Yet the sandwiched graphene is different.  The ``monolayer'' energy reference is preserved in the decoupled Dirac cone, causing a relative offset between the flat bands and the cone's vertex.
Comparison to experiments and fully self-consistent modeling is necessary to accurately assess $\Delta E_K$, yet the coexistence of flat bands and a weakly off-set Dirac cone at the magic angle is robust.

\textbf{Lattice relaxation effects}.--- Lattice relaxation effects are indispensable for understanding TSWG electronic structure at small twist angles, inducing renormalization of the quasiparticle spectrum near the Fermi energy and providing robust energetic stability to the flat bands.
An example of the relaxation patterns in TSWG is given in Fig. \ref{fig:relaxations}a.
The top and bottom layers (layers 1 and 3) are in $AA$ stacking and the middle layer (layer 2) is twisted $1.47^\circ$ counter clockwise relative to them.
The relaxation pattern is similar to that of twisted bilayer graphene: the relaxation fields form spirals around the $AA$ and $AB$ stacking regions, causing the effective twist angle between layers to change locally.
The spirals around $AA$ enhance the local twist angle, while the spirals around $AB$ reduce the local twist angle.
Overall, this maximizes the $AB$/$BA$ low energy stacking configuration and minimizes the area of the high-energy $AA$ stacking.
The displacements in layer 2 are the opposite sign and roughly twice the magnitude compared to layer 1 or 3.
This is expected as layer 2 experiences twice the interlayer potential of the other layers (see Methods).

The inclusion of atomic relaxation changes the graphene sandwich's electronic structure (see Fig. \ref{fig:relaxations}d).
The flattest bands occur at roughly $0.1^\circ$ smaller of a twist angle when compared to the unrelaxed case as shown in Fig. \ref{fig:bands_fig}c, and the gaps on both the electron and hole side of the flat bands at the $\Gamma$ point are significantly increased. 
This is similar to the effects of relaxation in TBG, and has to do with changes in the relative interlayer coupling strength between $AA$ and $AB$ stacked domains as well as the pseudo-gauge fields caused by in-plane strains.\cite{Carr2018Relaxation}

\textbf{Energetical stability, experimental viability}--- To address a  technical challenge of aligning external layers towards an $AA$ stacking, we investigate the effect of graphene layers misalignment.
In particular, we focus on the $\theta = 1.47^\circ$ graphene sandwich as a representative moir{\'e} supercell with the flat band regime (Fig. \ref{fig:relaxations}). For a trilayer graphene, one of the interlayer shifts can be safely eliminated by a corresponding shift of the reference frame; however there is still a remaining degree of freedom which alters the electronic band structure. To be concrete, we consider the relative displacement $\vec d_{13}$ of layer 3 with respect to the layer 1, where $\vec d_{13} = 0$  is a desirable condition for coexistence of flat bands with Dirac cones. 
We find that after shifting layer 3 with respect to layer 1 ($\vec d_{13} \ne 0$), the relaxation pattern does change, while the overall optimized energy does not.
In fact, the TSWG system after relaxations  behaves in the way that layers 1 and 3 translate to remove the initial displacement away from $AA$ stacking (Fig. \ref{fig:relaxations}a).
We remark that this phenomenon may be general for multilayered van der Waals structures, meaning stacking misalignment will not occur easily in fabricated devices.

To confirm that the  $AA$  stacking of the bread layers is robust against interlayer shifts, we fix the displacement of each layer to prevent translation back to $AA$ stacking (Fig. \ref{fig:relaxations}b) and calculate the total energy as a function relative displacement between layers 1 and 3 (Fig. \ref{fig:relaxations}c)
The $AA$ stacking order has an energy barrier of 20\,meV/nm$^2$ when the displacement between layers $1$ and $3$, labeled $\mathbf{d}_{13}$, is fixed. 
To compare with important energy scales, we use our model for twisted bilayer graphene which gives an energy difference of 30\,meV/nm$^2$ between relaxed and unrelaxed TBG at $\theta = 1^\circ$, and produces relaxation patterns in a good agreement with those observed in diffraction experiments.\cite{Yoo2019}
We thus conclude that the TSWG relaxation barrier of 20 meV/nm$^2$ is experimentally vital, and we expect that relaxation will cause graphene sandwich devices to naturally align the bread layers as $AA$, circumventing the experimental challenge of aligning the top and bottom layers manually.
We note that the $AA$ stacked alignment is the highest symmetry configuration of the system, yielding a 3-fold rotation ($D_3$) center.
This is a natural result, as most crystals minimize their energy by maximizing internal symmetry, except in some more exotic situations -- such as Peierls distortions and charge density waves.

\begin{figure}
	\centering{
\includegraphics[width=1 \linewidth]{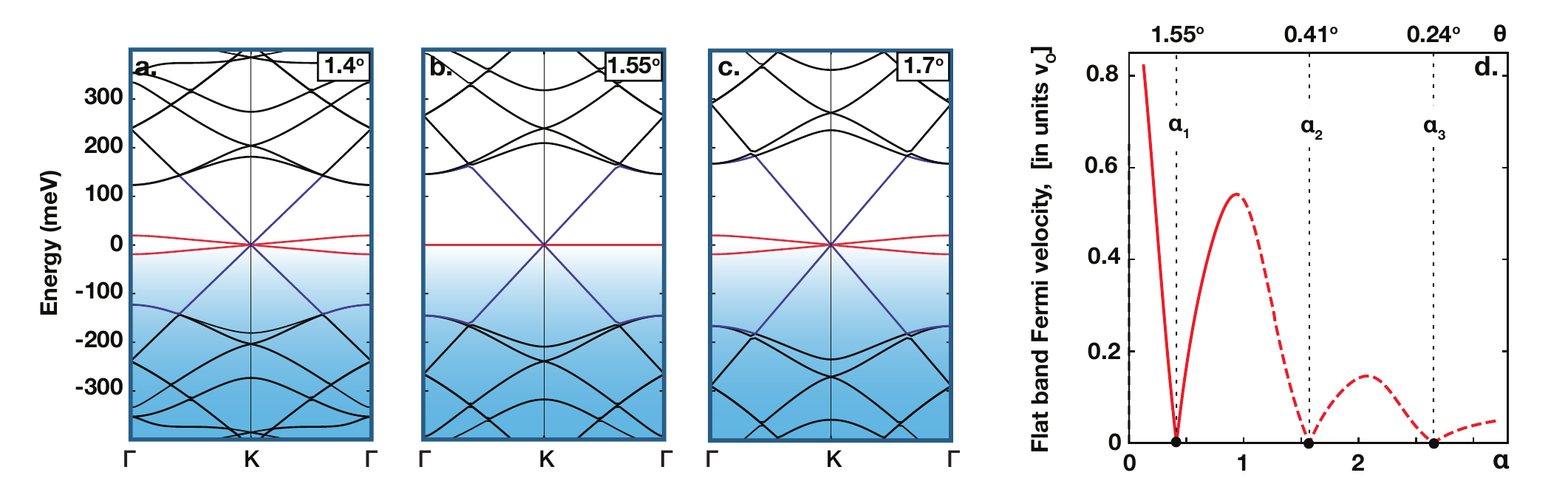}}	\caption{\footnotesize \textbf{Coexistence of ultraheavy and ultrarelativistic quasiparticles in idealized sandwiched graphene.} \textbf{(a)}-\textbf{(c)} Band structures for twisted graphene sandwich  in the continuum chirally-symmetric model below \textbf{(a)} and above \textbf{(c)} the magic angle condition \textbf{(b)}.  Exactly at the magic angle \textbf{(b)}, the low-energy quasiparticle spectrum is represented by flawlessly dispersionless bands piercing the steep Dirac cones through vertexes. Similar to the case of twisted bilayer graphene, the flat bands become dispersive both above \textbf{(a)} and below \textbf{(c)} the magic angle  tuning, showing evolution of the flat band Fermi velocities at moir{\'e} Dirac points with further twists \textbf{(d)}. The renormalization shown with the dashed line is likely affected by atomic relaxations not accounted here.
		\label{fig:cont_fig} }
\end{figure}

\textbf{Perfectly flat bands piercing Dirac cones}.--- The behaviour of flat bands piercing the Dirac cone can be most clearly understood in the chirally-symmetric limit of the effective continuum model (see Methods). At the magic angle twist 1.5$^{\circ}$, the moir{\'e} superlattice is approximately 40 graphene unit cells, and the effective behaviour of electrons in TSWG is governed by a large-period moir{\'e} field built on the three symmetry-related wave vectors $|\vec q_i|  = 2 k_D \sin ({\theta}/{2})$. For the given input monolayer Fermi velocity $v_0$, the continuum model for TSWG is captured by three key parameters: interlayer couplings between $AA$ and $AB$ sites ($w_{AA}$ and $w_{AB}$) as well as by the twist degree of freedom ($\theta$). Out-of-plane lattice relaxation affects the relative ratio of $w_{AA}/w_{AB}$, which is strongly suppressed at small angles justifying the use of a chiral-symmetric model with $w_{AA} =0$. Importantly, this model is ruled entirely by the only dimensionless twisting parameter $\alpha (\theta) = {w_{AB}}/{2 k_D v_0 \sin \frac{\theta}{2}}$, and shows the perfectly flat bands piercing the Dirac cone at $\alpha_* \simeq 0.414$ which for $w_{AB} = 110$\,meV corresponds to the magic angle $\theta_* \approx 1.55^{\circ}$  (see Fig. \ref{fig:cont_fig}).  
 The principle magic angle of TSWG, being exactly defined in the continuum model, is nearly $40\% $ larger than the reported magic angle in TBG, which makes TSWG experimentally attractive.

Fig.\ref{fig:cont_fig} shows the band structures for three instances of TSWG with twist angles exactly at the principal magic angle ($\alpha_* = 0.414..$), and just above and below this angle. We observe the perfectly flat bands piercing the Dirac cone vertex at $\theta = 1.55^{\circ}$ (Fig. \ref{fig:cont_fig}b), while for slightly different angles (Fig. \ref{fig:cont_fig}a,c) the bands are dispersive. We further track the renormalized Fermi velocities for TSWG, one from the Dirac cone itself and one from the flat bands. We define the  first Fermi velocity as the slope  corresponding to the flattened bands at the $K$ point and the second Fermi velocity corresponding to the Dirac cones. The first  Fermi velocity vanishes at $\theta_1=1.55^\circ$ and then reappears (see Fig \ref{fig:cont_fig}b), thus showing the behaviour similar to the case of twisted bilayer graphene.\cite{Bistritzer2010,Tarnopolsky2018} On the contrary, the second Fermi velocity is constant in this model and equals to the monolayer value $v_0$, confirming that the Dirac cone is a robust feature of TSWG, and the second Fermi velocity is very weakly dependent on twist as is reflected in our rigorous atomistic calculations. The continuum model for the twisted graphene sandwich also predicts higher-order magic angles (e.g. $\alpha_2 \simeq 1.57$, and $\alpha_3 \simeq 2.65$, which corresponds to $\theta_2 \approx 0.405^{\circ}$ and $\theta_3 \approx 0.240^{\circ}$), but these have not been confirmed by \textit{ab initio} calculations and are likely suppressed by lattice relaxations as in the bilayer case.\cite{carr2019}  We remark that the flawlessly flat band TSWG model ($w_{AA} = 0$) captures the principle magic angle accurately ($1.55^{\circ}$ vs $1.47^{\circ}$ in atomistic calculations with relaxations), but stretches the energy scales. The inclusion of a realistic $w_{AA} \approx 90$ meV produces the energy scales similar to \textit{ab initio} band structures, preserving the magic angle value. 

The continuum model is also interesting because the flat band condition can be derived \textit{analytically} up to an arbitrary precision. The leading order perturbation theory in $\alpha$ gives the following dependence of the flat band Fermi velocity on twist $\theta$:
\begin{equation}
v (\theta) = v_0 \,  \frac{ 1 - 6 \alpha^2(\theta)}{1 + 6 \alpha^2(\theta)}. 
\end{equation}
\noindent
The principal magic angle is thus precisely defined when the two Dirac cones hybridize into the flat band with $v(\theta_*) = 0$, which has a solution at
\begin{equation}    
\alpha_* = \frac{1}{\sqrt{6}} \approx 0.408,
\ \ \  \  \
\theta_*  =   \frac{w_{AB}}{ \alpha_* k_D v_0}. 
\end{equation}
\noindent
This estimate is also valid  for a more realistic case of $w_{AA} \ne 0$. The chiral symmetric model with $w_{AA}=0$ also has an exact mapping to the twisted bilayer graphene\cite{Khalaf2019}, and thus, in continuum theory, the graphene sandwich can have an infinite set of magic angles related to the TBG magic angles by the factor $\sqrt{2}$. For our discussion, only the principal magic angle is relevant.

\textbf{Tunability under external electric fields}.--- An advantage of the graphene sandwich is the tunability under external fields, with which the energy off-set between the Dirac cone and the Fermi level can be precisely controlled.
We show this with application of external fields and uniform strains to the unrelaxed structure to understand how the position of the Dirac vertex could be tuned in experimental devices.
We find that uniform planar strain can smoothly tune $\Delta E_K$, with a larger (smaller) lattice giving a smaller (larger) gap, which happens essentially due to rescaling of the system's characteristic energy.
As the lattice expands, the nearest-neighbour bonding distances increase and the electronic couplings become weaker.
The Fermi velocity of the Dirac cone is directly proportional to this nearest-neighbour coupling energy, and as previously discussed, the Fermi velocity sets an overall energy scale for the twisted graphene systems.
Thus, as the bonds are made weaker or stronger, the energy scale becomes smaller or larger, which directly relates to the value of $\Delta E_K$.

Applying an external vertical electric displacement field can control the magnitude of $\Delta E_K$ and tune the amount of hybridization between the flat bands and the Dirac cone (Fig. \ref{fig:tuning_fig}).
As mentioned earlier, this hybridization is set to be zero in our simplified tight-binding model in the absence of  displacement electric field, but we expect some non-zero hybridization at zero field because of the weak electronic coupling between the bread layers. 
When the displacement field is stronger than 0.4\,V/nm, the low-energy band structure is no longer easily comparable to the zero field case, as the flat bands become very dispersive and higher-energy bands begin intersecting the original flat bands near the Fermi energy.
To compare the displacement field modeled here to experimental devices, the dielectric screening of the graphene layers and any encapsulating substrate should be considered.
Regardless, sufficiently weak external electric fields provide tunable control of the relative position of the Dirac cone vertex and the intensity of the  van Hove singularity associated with the flat bands.

\textbf{Discussion.}--- We first discuss the implication of our TSWG results for experiments. Most importantly, the magic angle in TSWG is 1.5$^{\circ}$, which is $40 \%$ larger than in the parent TBG heterostructure. In general, this is advantageous for two reasons: first, it is easier to fabricate a multilayer heterostructure at larger twist angles; second, a qualitative trend is that larger angles (smaller moir\'e patterns) generally correspond to a higher superconducting $T_C$. As an example, in TBG the magic angle is 1.1$^{\circ}$ and $T_C =1.7$ K, but when applying appropriate hydrostatic pressure the magic angle superconductivity can be observed at larger angle (1.27$^\circ$) and larger $T_C$ (3.5\,K).\cite{Yankowitz2018}
Another example is the twisted double bilayer graphene (TDBG) which has the flat band region around 1.25$^{\circ}$, and superconducting $T_C  = 3.5$\,K.\cite{Shen2019, Liu2019, Cao2019} 
Additionally, the twisted sandwich graphene can be fabricated from the single flake of graphene.
First, one makes the twisted bilayer graphene with a standard approach.\cite{Frisenda2018} At small angles this heterostructure relaxes on the moir\'e length scale, which will assist the deposition of a third layer at the correct $0^\circ$ alignment with the first layer.
As we have shown in Fig. \ref{fig:relaxations}(c), even if the third layer is not positioned perfectly, it will tend to relax towards the energy minimum with depth of order 20 eV/nm$^2$. We note that this energy barrier is relatively large, as it is the same order of magnitude as between unrelaxed and fully relaxed TBG. 

The flat bands in the graphene sandwich exhibit a strong van Hove singularity in the density of electron states (see Fig. \ref{fig:relaxations}d). If realized experimentally, this will promote strongly correlated electron states as the kinetic energy is suppressed. Therefore, we may expect the emergence of correlated insulation and unconventional superconductivity under fine tuning, similar to TBG and TDBG. One of the main result of this study is that although at first glance the TSWG seems challenging to fabricate, our \textit{ab initio} results predict that the system will \textit{always} relax towards the beneficial $AA$ stacking between the first and third layers, providing the proper atomic geometry for the remarkably flat bands.

Last but not least, TSWG hosts both the stable flat bands and a Dirac cone in close proximity to one another in the band structure. There are only a few systems with this property, such as the exotic Kondo Weyl Semimetals and some Kagom\'{e} lattice systems with relatively flat bands offset from Dirac cones. All these systems are difficult to realize experimentally. 
From this perspective, TSWG may open a feasible experimental path for realizing the coexistence of both strongly localized and ultramobile quasiparticles simultaneously, important for the ``steep band/flat band" scenario of superconductivity. As the flat bands can pierce the Dirac cone extremely close to its vertex, TSWG can also be viewed as a potential platform for reaching the so-called \textit{triple point} states, a fine tuning for which the low-energy physics is effectively described by the Dirac equation with pseudospin-1 (see also Refs.\cite{Ramires2019,Zhu2016triple}).
To conclude, the twisted sandwiched graphene represents a novel, experimentally feasible platform for a broad range of exotic electronic phenomena.

\section{Methods}

\textbf{Density functional method}.-- We use the VASP \cite{Kresse1993, Kresse1996a, Kresse1996b} implementation of density functional theory (DFT) to calculate electronic structure for untwisted bilayer and trilayer graphene systems.
The semi-local meta-GGA functional SCAN+rVV10 \cite{Peng2016} is used for its good performance in van der Waals materials and low computational cost.
Multiple calculations of these untwisted systems are performed, with the graphene layers shifted in-plane to accurately capture electronic and mechanical effects of different stacking orders.
The graphene sandwich (bilayer) systems have 6 (4) carbon atoms, and we include a vertical vacuum space of $20\,\textrm{\AA}$ to prevent interactions between periodic images in the $z$ direction of the heterostructures.
Then \textit{ab initio} tight-binding parameters are extracted by the method of maximally localized Wannier functions with the wannier90 package.
From this, one obtains a fully parameterized model for in-plane and inter-plane $p_z$ orbital interactions, which can also accurately account for corrugations and in-plane strain.\cite{Fang2016, Fang2018, Carr2018a,Carr2018Relaxation}
We find that the largest effective tight-binding coupling between $p_z$ orbitals of the top and bottom layers of the sandwich is $6$\,meV, roughly $2\%$ of the maximum coupling between adjacent layers.
For this reason, we ignore couplings between the top and bottom layers in tight-binding and continuum simulations.

\textbf{Lattice relaxations modelling}.--The relaxation of TSWG is obtained using a continuum model to account for in-plane distortions due to relaxation and the Generalized Stacking Fault Energy (GSFE) to account for the interlayer coupling.
The relaxation of the layer $i$, $\bm u_i (\bm b)$, is defined in terms of the local configuration or the relative local stacking,  $\bm b$, and we obtain $\bm u_i (\bm b)$'s by minimizing the total energy.\cite{Cazeauz2018} The energy has two contributions. The first is the intralayer energy of the $i$-th layer, which is calculated based on linear elasticity theory, 
\begin{align}
E_\mathrm{intra} [\bm u_i (\bm b)] 
 & =  \int_{\Gamma} \,d \bm b  \, \frac{1}{2} \mathcal{E}(\grad_{\bm b} \bm u_i (\bm u_i))\,  C_i \, \mathcal{E} (\grad_{\bm b} \bm u_i (\bm b)) \nonumber 
 \\
&  = \int_{\Gamma}  d \bm b \, \frac{1}{2} \Big[G (\partial_x u_{i,x} + \partial_y u_{i,y})^2    \nonumber \\
&   + K ( (\partial_x u_{i,x} - \partial_y u_{i, y})^2 + (\partial _x u_{i,y} + \partial_y u_{i,x})^2) \Big],
\end{align}
\noindent
where $\Gamma$ is the union of all configurations, $\mathcal{E}(\grad \bm{u}_i) $ is the strain tensor, $C_i$ is the linear elasticity tensor of the $i$-th layer (which is identical for all $i$'s in this case), $G$ and $K$ are shear and bulk modulus of a monolayer graphene, which we take to be $G = 47352 \, \mathrm{meV/cell}$, $K = 69518 \, \mathrm{meV/cell}$ and the graphene unit cell size is $5.3128\,$\AA$^2$. The values of $G$ and $K$ are obtained with DFT calculations by isotropically straining and compressing the monolayer and performing a linear fitting of the ground-state energy as a function of the applied strain. The second energy contribution is the interlayer energy, which is described by the GSFE.\cite{Carr2018Relaxation, Dai2016, Zhou2015} The GSFE, denoted as $V_\mathrm{GSFE} (\bm b)$, has been employed to explain relaxation in van der Waals heterostructures,\cite{Carr2018Relaxation} which depends on the relative stacking between two adjacent layers. We obtain the $V_\mathrm{GSFE} $ by applying a $9 \times 9$ grid sampling of rigid shifts to layer 1 in the unit cell with respect to layer 2 and extract the relaxed ground state energy at each shift from DFT. The GSFE of graphene at a given configuration $\bm b = \begin{pmatrix} v & w \end{pmatrix} ^T$ can then be expressed as follows, 
\begin{align}
V_\mathrm{GSFE} (v, w) &  = c_0 +  c_1(\cos v + \cos w + \cos (v + w) ) \nonumber \\
& +  c_2 (\cos (v + 2w) + \cos(v-w) + \cos(2 v + w)) \nonumber \\
& +  c_3 (\cos 2 v  + \cos 2 w + \cos(2 v + 2 w)), 
\end{align}
\noindent
where $c_i$'s are coefficients found by fitting the ground state energy at each shift. $c_0 = 6.832 \, \mathrm{meV/cell}$, $c_1 = 4.064 \, \mathrm{meV/cell}$, $c_2 = -0.374 \, \mathrm{meV/cell}$, $c_3 = -0.095 \, \mathrm{meV/cell}$. In terms of the $V_\mathrm{GSFE}$, the total interlayer energy $E_\mathrm{inter}$ can be then written as follows for a relaxed TSWG, 
\begin{equation}
E_\mathrm{inter} [\bm u_1, \bm u_2] = 2 \int_\mathrm{\Gamma} d \bm b \, V_\mathrm{GSFE} (\bm b + \bm u_1 (\bm b) -  \bm u_2 (\bm b)) , 
\end{equation} 
\noindent
where the factor of 2 comes from the sum of couplings between layers 1, 2 and layers 2, 3, and we use the fact that $\bm u_1 (\bm b) = \bm u_3 (\bm b)$ due to the layer inversion symmetry in the TSWG system.  Note that the $V_\mathrm{GSFE}$ is a function of the sum of unrelaxed configuration and the relaxation displacement vectors in order to describe the inter-layer stacking energy after relaxation. The total energy is the sum of the interlayer and the intralayer energies: 
\begin{equation}
E_\mathrm{tot} (\bm u (\bm r)) = \sum_{i=1}^3 E_\mathrm{intra} (\bm u_i (\bm b)) + E_\mathrm{inter} ( \bm u_1 (\bm b), \bm u_2 (\bm b)).
\end{equation}
The relaxation $u_i (\bm b)$ is computed by minimizing the total energy. The following linear transformation maps the relaxation from the local configuration to the real space positions  $\bm r$:
\begin{equation}
\bm b = (E_1^{-1} E_2 - \mathbb{1}) \bm r, 
\end{equation}
where $E_1$ and $E_2$ are the unit cell vectors of the first (unrotated) and the second layers (rotated counter clockwise by $\theta$), respectively.

\textbf{Tight binding calculations}.--We take $a_1 = \alpha (1, 0)$ and $a_2 = \alpha (\frac{1}{2}, \frac{\sqrt{3}}{2})$ with $\alpha = 2.4768 \textrm{\AA}$ as the unit cell vectors of graphene.
Periodic supercells are constructed in terms of the integers $n$ and $m$ according to $\cos\theta = \frac{n^2 + 4nm + m^2}{2(n^2 + nm + m^2)}$.
For the trilayer sandwich, the bottom and top layers are untwisted while the middle layer is twisted counterclockwise by $\theta$. 
This gives supercell vectors of $(n a_1 - m a_2)$ and $(-m a_1 + (m+n) a_2)$ for the untwisted layers
and supercell vectors of $(m a'_1 - n a'_2)$ and $(-n a'_1 + (m+n) a'_2)$ for the twisted layer ($a'_i$ are $a_i$ rotated c.c.w by $\theta$).
Band structures are calculated by diagonalization of these supercell Hamiltonians.
The interlayer electronic couplings deal with relaxation and strain effects easily, as they depend directly on the $p_z$ orbital positions by design.
Strain is included for in-plane coupling with a simple bond-length approximation, where the coupling $t$ is dependent on the bond-length $r$ by $t = t_0 + \alpha_t \frac{r-r_0}{r_0}$.\cite{Fang2018}
The vertical electric displacement field effect is treated in the leading order by adding on-site energies $\delta = E_z r_z$  in the existing tight-binding model.

\textbf{Continuum model.} We use an effective  continuum model based on approach of Refs\cite{Tarnopolsky2018,Khalaf2019}.  In the pristine setting, the effective continuum model for the twisted graphene sandwich  can be described by
\begin{align}
\label{Htrilayer}
\mathcal{H} = 	
\begin{pmatrix}
- i \hbar v_0 ( \boldsymbol \sigma_{-\theta/2} \cdot \boldsymbol{\nabla} )  & T (\vec r) & 0 
\\
T^{\dag} (\vec r)	& - i \hbar v_0 ( \boldsymbol \sigma_{+\theta/2}  \cdot \boldsymbol{\nabla}) & T^{\dag}(\vec r) 
\\
0 & T(\vec r) & - i \hbar v_0 ( \boldsymbol \sigma_{-\theta/2} \cdot \boldsymbol{\nabla} )
\end{pmatrix},
\end{align}
\noindent
where the moir{\'e}-induced interlayer coupling is taken up to the first shell in momentum space,
\begin{align}
T(\vec r)  = \sum_{n =1,2,3}
T_n \, e^{- i \vec q_n \vec r}, 
\nonumber
\end{align}
\noindent
with the three-fold star of $\vec q_i$,  $|\vec q_j| = 2 k_D \sin (\theta/2)$, each equirotated by $\phi = 2 \pi/3$, and  
\begin{align}
T_n  & = e^{- i \boldsymbol{\mathcal{G}}^{(n)}_{\theta} \vec d} \ 
\hat \Omega_{\phi}^{n-1}  
\begin{pmatrix}
w_{AA} & w_{AB} 
\\
w_{AB} & w_{AA}  	
\end{pmatrix}
\hat \Omega_{\phi}^{1-n},
\nonumber   	
\end{align}
\noindent
where $\boldsymbol{\mathcal{G}}^{(0)}_{\theta} = 0$, $\boldsymbol{\mathcal{G}}^{(1)}_{\theta} = \vec q_2 - \vec q_1$, $\boldsymbol{\mathcal{G}}^{(2)}_{\theta} = \vec q_3 - \vec q_1$ are the moir{\'e} reciprocal cell vectors  and $\vec d$ is the relative displacements of one layer with respect to another one, and
\begin{align}
\hat \Omega_{\phi}  & = 
\begin{pmatrix}
0 & e^{+ i \phi} 
\\
e^{- i \phi}  & 0  	
\end{pmatrix}.
\nonumber
\end{align}
\noindent
Note that in the continuum model, one can eliminate one of the displacements after redefining the reference frame. The Hamiltonian of Eqn. \eqref{Htrilayer}  acquires additional chiral symmetry and perfectly flat bands piercing Dirac cones at neutrality for $w_{AA} = 0$.  In this idealistic setting, the twisted trilayer graphene have a family of well-defined  magic angles, enlarged compared to the similar sequence in the TBG by $\sqrt{2}$.  See Ref.\cite{Khalaf2019} for further details.

\begin{addendum}
 \item We thank Pablo Jarillo-Herrero, Bertrand Halperin and Oleg Yazyev for useful discussions.  This work was performed under Grant No.  P2ELP2\_175278 from the Swiss National Science Foundation (SNSF), Grants No. DMR-1664842 and No. DMR-1231319 from the National Science Foundation (NSF), and ARO MURI Award W911NF-14-0247.
The computations  were run on the Odyssey cluster supported by the FAS Division of Science, Research Computing Group at Harvard University. 

 \item[Competing Interests] The authors declare  no
competing financial interests.
 

\end{addendum}

\bibliography{Refs.bib}

\end{document}